\documentclass[
 aip,
 amsmath,amssymb,
 reprint,%
]{revtex4-1}

\usepackage{graphicx}
\usepackage{dcolumn}
\usepackage{bm}

\usepackage{comment}
\usepackage[utf8]{inputenc}
\usepackage[T1]{fontenc}
\usepackage{tikz}
\usepackage{mathptmx}
\usepackage{etoolbox}
\usepackage{braket}
\usepackage{subcaption}
\usepackage{tabularx}
\usepackage{adjustbox}
\usepackage{multirow} 

\makeatletter
\def\@email#1#2{%
 \endgroup
 \patchcmd{\titleblock@produce}
  {\frontmatter@RRAPformat}
  {\frontmatter@RRAPformat{\produce@RRAP{*#1\href{mailto:#2}{#2}}}\frontmatter@RRAPformat}
  {}{}
}%
\makeatother

\begin{document}

\preprint{AIP/123-QED}

\title[Design of Magnetic Lattices with {a Quantum-Inspired Evolutionary} Optimization Algorithm]{Design of Magnetic Lattices with {a Quantum-Inspired Evolutionary} Optimization Algorithm}
\author{Zekeriya Ender E\u{g}er}
\affiliation{ 
Virginia Tech, Blacksburg 24060 VA, USA; email: endereger@vt.edu
}%
\author{Waris Khan}
\affiliation{ 
Virginia Tech, Blacksburg 24060 VA, USA; email: warisk@vt.edu
}%
\author{Priyabrata Maharana}
\affiliation{BosonQ Psi (BQP), Syracuse 13202 NY; email: priyabrata.m@bqpsim.com}
\author{Kandula Eswara Sai Kumar}
\affiliation{BosonQ Psi (BQP), Syracuse 13202 NY; email: eswara.sai@bqpsim.com}
\author{Udbhav Sharma}
\affiliation{BosonQ Psi (BQP), Syracuse 13202 NY; email: udbhav.sharma@bqpsim.com}
\author{Abhishek Chopra}
\affiliation{BosonQ Psi (BQP), Syracuse 13202 NY, USA; email: abhishekchopra@bqpsim.com}
\author{Rut Lineswala}
\affiliation{BosonQ Psi (BQP), Syracuse 13202 NY, USA; email: rutlineswala@bqpsim.com}
\author{P{\i}nar Acar*}%
 \email{pacar@vt.edu}
\affiliation{ 
Virginia Tech, Blacksburg 24060 VA, USA
}%


\date{\today}

\begin{abstract}
This article investigates the identification of magnetic spin distributions in ferromagnetic materials by minimizing the system’s free energy. Magnetic lattices of varying sizes are constructed, and the free energy is computed using an Ising model that accounts for spin-to-spin neighbor interactions and the influence of an external magnetic field. The problem reduces to determining the state of each spin, either up or down, leading to an optimization problem with $2^{n \times n}$ design variables for a $n \times n$ lattice. To address the high-dimensional and computationally intractable nature of this problem, particularly for large domains, we employ a quantum{-inspired evolutionary} optimization {(QIEO)} algorithm, part of the  {BQPhy\textregistered} {QuantumNOW\texttrademark} solver. QIEO results are first validated against solutions obtained using a genetic algorithm for smaller lattices. Finally, the approach is extended to large-scale systems, including $50 \times 50$ lattices, where conventional methods become impractical.

\end{abstract}

\maketitle


\section{\label{sec:level1}Introduction}

Magnetic materials continue to captivate researchers and engineers due to their critical role in numerous technological applications, particularly in sectors such as energy and transportation. Understanding and determining the magnetic behavior on small length scales is key to maximizing functional stability and magnetic efficiency under varying operational conditions.

To model and analyze the complex behavior of magnetic systems, a variety of theoretical frameworks have been developed. These models broadly fall into two categories: those based on localized magnetic moments and those described by electronic band structures. Among the latter, the Hubbard \cite{herrmann1997magnetism}, Stoner \cite{tannous2008stoner}, and Kondo \cite{fazekas1991magnetic} models, as well as approaches rooted in Density Functional Theory (DFT) \cite{bihlmayer2020density}, provide insight into the role of electronic structure and correlations. In contrast, models such as the Heisenberg and Ising frameworks deal with discrete spin variables and their interactions, with the Ising model \cite{ising1925beitrag} being particularly effective in representing simplified spin systems.

In this study, the Ising model is employed to explore the landscape of spin configurations and to determine those that minimize or optimize the system's free energy. As with other Hamiltonian-based models, the Ising framework calculates the total energy of a system as a function of spin arrangements and their interactions, serving as a foundational approach in statistical mechanics \cite{cipra1987introduction}. Despite its usefulness, the model's computational demands scale rapidly with lattice size and dimensionality, posing challenges for classical simulations. These limitations have spurred significant research on numerical efficiency, including efforts to address scaling behavior, spin correlation functions, and hysteresis effects \cite{sethna2004random}. This causes limitations on investigating the effects of the uncertainty arising from external parameters, including the temperature and external magnetic fields. It is also the reason for potentially numerically intractable solutions when sampling-based uncertainty quantification (UQ) algorithms (e.g., Monte Carlo Simulation (MCS)) are incorporated to investigate the effects of uncertainty on these systems. As solving the optimal spin configurations to determine the magnetic state of the materials under the effects of uncertainty requires a computationally demanding and expensive \textit{design-under-uncertainty} solution, there is a need to shift from more traditional optimization solutions (e.g., genetic algorithm) that can address the discrete nature of this problem to more advanced computing strategies.



{To overcome these challenges, optimization methods that employ quantum principles have emerged as promising alternatives. In particular, Quantum-Inspired Evolutionary Optimization (QIEO) combines classical evolutionary strategies with computational structures inspired by quantum mechanics, most notably the concepts of quantum superposition and rotation gates \cite{kumar2024qieo}. 
This foundation was built on earlier work by Han and Kim \cite{han2000genetic}, whose Quantum-Inspired Evolutionary Algorithm (QEA) demonstrated that Q-bit representations and quantum gate updates could efficiently navigate discrete optimization landscapes even with small population sizes. QEA also highlighted the role of quantum observation in collapsing probabilistic states into deterministic solutions and demonstrated robust performance across several combinatorial problems, including the 0–1 knapsack and flow-shop scheduling problems \cite{QIEOReview}, which motivated subsequent extensions to parallel formulations \cite{han2001parallel}. Collectively, these developments established the broader QIEO framework, which has since proven effective for high-dimensional and highly constrained discrete design tasks, such as those found in magnetic spin lattice design \cite{han2002quantum, rylander2004quantum, zhang2006novel}.}

A closely related approach is the Quantum Approximate Optimization Algorithm (QAOA) \cite{farhi2014quantum}, which formulates discrete optimization problems as minimizing the expectation value of a problem-specific Hamiltonian on quantum states parameterized by a variational circuit. Although QAOA is typically run on quantum hardware, its mathematical foundation shares key principles with QIEO, notably, encoding solutions in the state vector of a quantum system and evolving them via parameterized unitaries that are analogous to update steps in QIEO. Moreover, as demonstrated by Lloyd et al. \cite{lloyd2016quantum}, quantum algorithms have the potential to efficiently capture topological and geometric features in high-dimensional configuration spaces. This is particularly valuable in problems like spin lattice optimization, where the solution landscape is highly nonconvex and spatial correlations between variables (i.e., spins) play a critical role in defining system energy. These insights reinforce the applicability of {BQPhy's} hybrid classical-quantum strategy in navigating complex magnetic design spaces with embedded physical constraints \cite{kumar2024qieo, kumar2025qieo}.



This work builds upon these insights by utilizing quantum-enhanced methods to directly identify optimal spin configurations, shifting the emphasis away from transition temperatures and toward configuration-space optimization as a central goal in magnetic material analysis.

\section{Ising Model Formulation}

\subsection{Mathematical Formulation}
The Ising model describes interactions among spins, which represent the atomic magnetic moments arranged on a lattice. Beyond condensed matter physics, it has found wide applications in diverse fields, including sociology and neural networks.\cite{sornette2014physics}. In this study, the Ising model is used to investigate the magnetic behavior of 2D lattices while incorporating short- and long-range order interactions between the spins.

The energy function for the Ising model is defined as~\cite{selinger2016introduction}:

\begin{equation}
    H=-\omega_{j}\Sigma_{<i,j>}\sigma_{i}\sigma_{j}-h\Sigma_{i}\sigma_{i}
    \label{eq:ener}
\end{equation}

In Eq. \ref{eq:ener}, $h$ is the external magnetic field, $\sigma$ shows the spin parameter, while the $\sigma_{i}\sigma_{j}$ term represents interactions between spins indexed by $i$ and $j$. Moreover, $\omega_j$ is the weight parameter defining the weighted importance of spin-to-spin interactions based on distance or order of neighboring information.  Accordingly, this weight parameter correlates with the degree of neighboring information defined in the Hamiltonian energy equation.

Solving the magnetic state of the material requires the minimization of the energy function. However, when large lattices are considered, the computation of the minimum free energy requires many iterations that are potentially intractable. To address this, it is common to only consider the interactions within the nearest neighbor spins. Nevertheless, only considering the nearest-neighbor interactions has some drawbacks, as long-range interactions can play an effective role in determining the magnetic state of the material. As a result, the modeling with only nearest-neighbor interactions may overlook important contributions from long-range neighbors or external factors such as external magnetic fields. To improve the prediction accuracy for the magnetic state of the material, the long-range order interactions are accounted for by implementing a Gaussian weight formulation as discussed in the next section.

\subsection{\label{sec:level2} Investigation of Spin-to-Spin Interactions}

Spin-to-spin correlation functions quantify how the magnetic orientation (spin) of a particle relates to that of another at a spatially distinct site within a material. These correlations provide critical insight into the extent and nature of magnetic interactions across a system. The characteristic length scale over which spins remain correlated, which is defined as the spin correlation length, serves as an effective order parameter. It is particularly useful in capturing the emergence or breakdown of long-range magnetic order \cite{stanley1971introduction,yeomans1992statistical}. Near critical points, such as those associated with phase transitions, this length can diverge, marking significant structural or dynamical changes in the spin configuration \cite{kardar2007statistical}. Variations in external parameters, including temperature and applied magnetic fields, modulate these correlations and give rise to distinct magnetic phases and critical behavior. Thus, a detailed understanding of spin correlations is not only fundamental for probing the onset of phase transitions but also essential for predicting the response of magnetic systems to external perturbations.

To incorporate long-range interactions within the spin system, the interaction strength, denoted by $\omega_j$, is characterized through a Gaussian distribution \cite{hasan2022uncertainty}. The Gaussian formulation used to describe the spatial variation of interaction strength is presented in Eq.~\ref{eq:mdl}, where $d$ represents the inter-spin distance, and $\mu_d$ and $\sigma_d$ correspond to the mean and standard deviation of the distance distribution, respectively \cite{euger2024uncertainty,eger2024uncertainty}. Furthermore, $\mu_d$ and $\sigma_d$ are defined as temperature-dependent variables modeled by additional Gaussian functions, such that $\mu_d = \mu_d(T)$ and $\sigma_d = \sigma_d(T)$, enabling the model to capture thermally driven changes in the correlation profile across the spin lattice \cite{euger2024uncertainty}.

\begin{equation}
    \omega=f(d,\mu_d(T),\sigma^2_d(T))=\frac{1}{\sqrt{2\sigma^2\pi}}e^{-(d-\mu)^2/2\sigma^2}\label{eq:mdl}
\end{equation}

At low temperatures, the interaction strength between spins tends to be nearly uniform, regardless of the spatial separation between them. However, as the temperature increases, this distribution shifts, which results in a diminished influence from distant neighbors relative to those in closer proximity. This attenuation of long-range interaction strength persists even beyond the phase transition, ultimately leading to a generalized weakening of spin interactions across the lattice, including those between nearest neighbors.

While the precise functional form of interaction strength across all temperature values is not explicitly defined in the literature, the behavior of the system is well-characterized by spin correlation functions under three primary regimes: above, below, and at the Curie temperature ($T_c$), which marks the critical point for the ferromagnetic to paramagnetic phase transition \cite{wipf2013statistical}. To effectively incorporate long-range interactions in our model, we construct a statistical formulation constrained to capture this behavior accurately across temperature regimes.

\begin{equation}
    \omega_{n} \leq \omega_{n-1} \leq \omega_{n-2} \leq....\leq \omega_{3}\leq \omega_{2} \leq \omega_{1}
\end{equation}

Here, the parameter $n$ is determined by the lattice size, while $\omega_{1}$ and $\omega_{n}$ represent the interaction strengths of the nearest and farthest neighboring spins, respectively. To account for long-range magnetic interactions, a window parameter $\omega_{n}$ is introduced. This parameter, which scales with the lattice size, ensures that all spin interactions across the lattice are captured. The window-based formulation has previously proven effective in predicting the spatio-temporal evolution of microstructures within an Ising model framework using Markov Random Fields \cite{acar2019epistemic,acar2016markov}, and has also been successfully employed in Ising models to investigate the onset of the ferromagnetic–paramagnetic phase transition \cite{euger2024uncertainty}.

\subsection{Modeling the Uncertainty of Temperature and External Field}

In realistic settings, temperature and external magnetic field values often exhibit uncertainty due to factors such as sensor inaccuracies, environmental disturbances, and system-level noise. To account for these sources of variability, both temperature ($T$) and magnetic field strength ($h$) are modeled as independent random variables by assuming that they agree with a Gaussian distribution with the following mean values: $T_\text{mean} = 298\text{K}$ and $h_\text{mean} = 5$, with each parameter subject to a standard deviation equal to 10\% of its respective mean value, capturing expected deviations in experimental or operational contexts.

To efficiently sample the input space defined by these random parameters, Latin Hypercube Sampling (LHS) is employed. A total of 3,000 design samples are generated to ensure broad and stratified coverage of the input domain. Following the sampling phase, the genetic algorithm is implemented to identify the optimum spin configurations under uncertainty. The genetic algorithm is well-suited to this task due to its effectiveness in exploring high-dimensional, non-convex solution spaces, particularly in the presence of complex spin interactions and long-range correlations. The optimization problem is further elaborated in the next section.

\section{Optimization of Magnetic Spin States under Uncertainty}

The states of magnetic spins are identified while accounting for the uncertainty of temperature and magnetic field strength. The stochastic optimization problems are defined using two different objectives. The first optimization problem aims to solve for the optimum {spin state $m$} that minimize the expected value of the system's free energy ($E[f_m]$) to detect the magnetic equilibrium state, while the second problem also considers the effects of the uncertainty of temperature and external magnetic fields by minimizing the resultant deviations in the free energy value ($std_{f_m}$). The problems are first solved using the genetic algorithm. By considering the computationally expensive nature of this global optimization algorithm, the problems are solved in a $10 \times 10$ magnetic domain. The problems are formulated as shown in Eqs. \ref{eq:first_prob} and \ref{eq:second_prob}, respectively:

\begin{equation}
\min E[f_m] 
\label{eq:first_prob}
\end{equation}
\begin{equation*}
\text{find      } \sigma_{i, j} \ \ \ \text{where     } i, j = 1, 2,..., 10
\end{equation*}

\begin{equation}
\min std_{f_m} 
\label{eq:second_prob}
\end{equation}
\begin{equation*}
\text{find      } \sigma_{i, j} \ \ \ \text{where     } i, j = 1, 2,..., 10
\end{equation*}

\begin{figure}[h!]
\centering
    \includegraphics[width=0.25\textwidth]{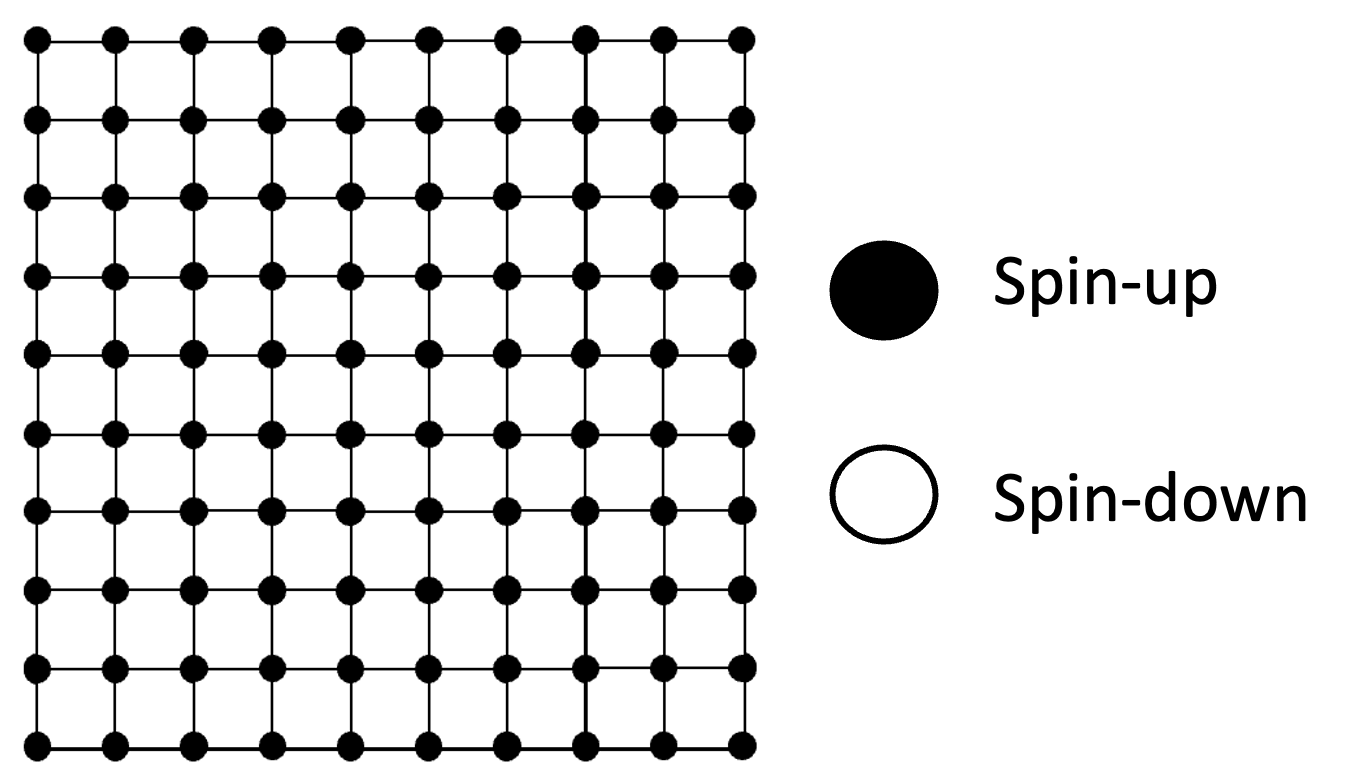}
\caption{Solution of the first optimization problem identifying the optimum spin distributions that minimize the expected value of the system's free magnetic energy under the effects of the uncertainty of temperature and external magnetic field using the genetic algorithm.}
\label{fig:problem1}
\end{figure}
The optimum solution for the first problem is obtained within just a few seconds as shown in Fig. \ref{fig:problem1}, so no additional optimization algorithm is required. In contrast, uncertainty significantly impacts the optimum solution for the second problem, unlike the first problem that yields the same result as the deterministic case. Because uncertainty influences the optimum design, classical algorithms such as GA require considerably more computational time. This highlights the need for a more efficient optimization algorithm.

\begin{figure}[h!]
\centering
    \includegraphics[width=0.5\textwidth]{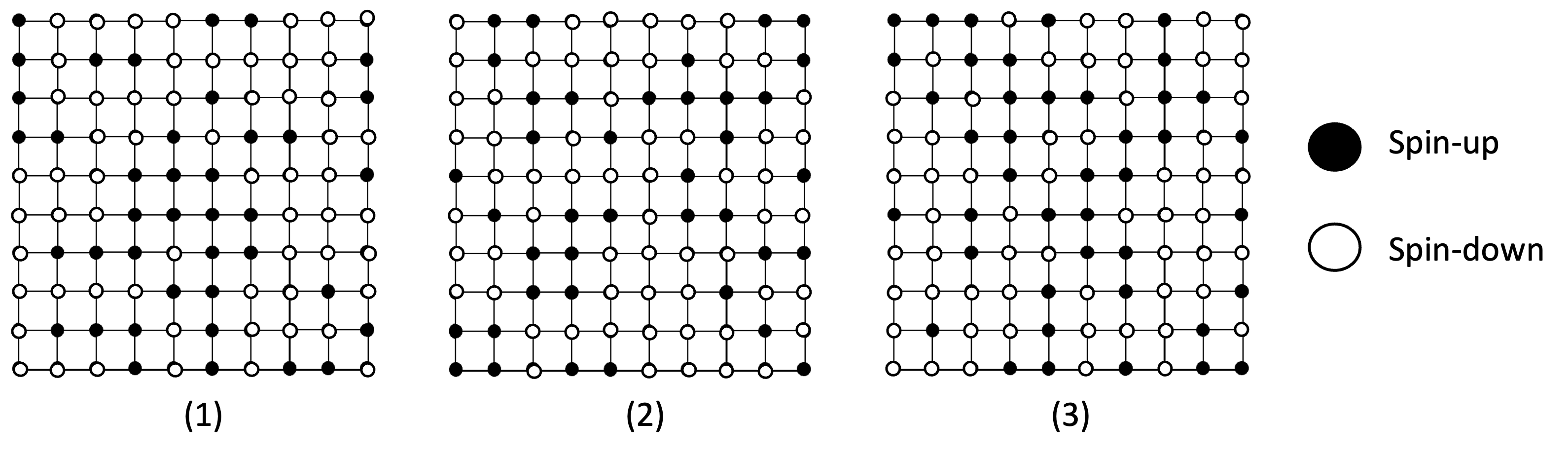}
\caption{Exemplary solutions of the second optimization problem identifying the optimum spin distributions that minimize the standard deviation of the magnetic free energy under the effects of the uncertainty of temperature and external magnetic field using the genetic algorithm.}
\label{fig:problem3}
\end{figure}


The second problem is also found to lead to multiple optimum solutions, all providing the same value for the objective function, and the reported times for the genetic algorithm and BQP include the solution time for any solution shown in Fig. \ref{fig:problem3}.

\section{Application of Quantum Optimization Algorithm}
\subsection{Quantum-Inspired Evolutionary Optimization Algorithm}
BQPhy QuantumNOW ({Q-NOW\texttrademark}) optimization solver is primarily a population-based metaheuristic algorithm that uses the quantum-inspired evolutionary optimization (QIEO) algorithm. QIEO utilizes traditional HPCs and is developed based on the principles of quantum computing. It primarily uses concepts such as qubits, quantum gates, superposition, and entanglement to efficiently balance exploration and exploitation. Unlike gradient methods, QIEO is a global search algorithm that starts with initial random solutions, each of which mimics a quantum circuit. These circuits are being updated using parametric gates based on the employed heuristics. The circuit update continues until it converges, i.e., reaches the optimal solution. 

{The basic unit of information in a quantum computer is the quantum bit, or qubit. Unlike a classical bit that can only be in the states $|0\rangle$ or $|1\rangle$, a qubit can exist in a superposition of both. The general state of a qubit is
\begin{equation}
|\psi\rangle = \alpha |0\rangle + \beta |1\rangle,
\label{eq: qubit definition}
\end{equation}
where $\alpha$ and $\beta$ are complex probability amplitudes satisfying $|\alpha|^2 + |\beta|^2 = 1$. This state can also be written in vector form as $\begin{bmatrix}\alpha \\ \beta\end{bmatrix}$.}

{A system of $m$ qubits can exist in a superposition of $2^m$ classical states:
\begin{equation}
\ket{\psi}^{\otimes m} = \sum_{\mathbf{x} \in \{0,1\}^m} p_\mathbf{x} \ket{\mathbf{x}}
\end{equation}
where $\mathbf{x}$ denotes an $m$-bit classical configuration and $p_\mathbf{x}$ is its probability amplitude. A convenient representation of an $m$-qubit quantum individual is
\begin{equation}
\mathbf{q} =
\begin{bmatrix}
\alpha_1 & \alpha_2 & \dots & \alpha_m \\
\beta_1 & \beta_2 & \dots & \beta_m
\end{bmatrix}.
\end{equation}
In evolutionary optimization, each classical state $\mathbf{x}$ corresponds to a point in the design variable space. Thus, a quantum individual $\mathbf{q}$ represents a superposition of $2^m$ such points, and increasing $m$ improves the resolution of the design space. A collection of quantum individuals forms a quantum population $ Q = \{q_1, q_2, \dots, q_n\}$.}

{To initialize the population, the Hadamard gate is applied to each individual to transit it to the uniform superposition state
\begin{equation}
H^{\otimes m}\ket{0^m} =  \sum_{\mathbf{x} \in \{0,1\}^m} \frac{1}{\sqrt{2^m}}\ket{\mathbf{x}},     
\end{equation}
which distributes equal probability amplitude across all $2^{m}$ classical basis states. In Step 5, we need to prescribe a parameter theta ($ \theta$), which helps to update the chromosomes and affects the exploration and the exploitation rate of the solution. This updates populations through the quantum rotation as expressed by}
{
\begin{equation}
\begin{bmatrix}
\alpha ^ {t} \\
\beta ^ {t}\\
\end{bmatrix} = \begin{bmatrix}
\cos(\theta) & -\sin(\theta) \\
\sin(\theta) &  \cos(\theta)
\end{bmatrix}
\begin{bmatrix}
\alpha ^ {t-1} \\
\beta ^ {t-1}\\
\end{bmatrix}
\end{equation} 
}
{where $\begin{bmatrix}
\alpha ^ {t-1} \\
\beta ^ {t-1}\\
\end{bmatrix}$ and $\begin{bmatrix}
\alpha ^ {t} \\
\beta ^ {t}\\\end{bmatrix}$ are the amplitudes of the $j^{th}$ quantum individual's $i^{th}$ qubit before and after the updating, respectively. This rotation gate ($R_{Y}$) modifies probability amplitudes in the direction of solutions with better fitness. In doing so, QIEO incrementally reshapes the underlying probability landscape such that the likelihood of generating a global optimum increases over time}. 
The general workflow of BQPhy Q-NOW QIEO is as follows:
\begin{enumerate}
    \item Initialize qubit population: All qubits in equal superposition (e.g., 50-50 chance of 0 or 1).
    \item Sample individuals from qubit states.
    \item Evaluate the fitness of the individuals sampled.
    \item Select the best solutions.
    \item Update qubits using rotation rules for the best solutions.
    \item Repeat until the stopping condition is met (max generations, fitness threshold, etc.).
\end{enumerate}

In general, QIEO offers a reliable solution that surpasses its traditional counterpart in terms of quality and consistency. The fact that QIEO requires a lower population to achieve better accuracy makes this algorithm suitable for solving complex optimization problems \cite{kumar2024qieo,kumar2025qieo}. 
\subsection{Classical Benchmark Algorithms}
To evaluate the computational efficiency of the quantum-inspired optimization framework ({QIEO}), we compare its performance with a classical genetic algorithm approach (GA) as well as simulated annealing (SA), which is a widely used baseline for energy minimization of spin-based systems \cite{isakov2015optimised,SA2}, implemented on a high-performance personal workstation. Both algorithms used to minimize the free-energy landscape and identify optimal spin configurations in the presence of uncertainty.
\subsubsection{Genetic Algorithm}
Genetic algorithm is a population-based optimization method inspired by the principles of natural selection and biological evolution \cite{goldberg1989genetic}. A population of candidate solutions is evolved over multiple generations instead of improving a single solution. New candidate solutions are generated through genetic operations that are selection, crossover, and mutation. Selection chooses individuals based on their fitness, crossover combines information from parent solutions to create offspring, and mutation introduces small random variations to maintain diversity in the population. Through repeated generations of evaluation and reproduction, the population evolves toward improved solutions. GA is particularly useful for complex optimization problems with large or non-convex search spaces where gradient-based methods may not be applicable.
\subsubsection{Simulated Annealing}
Simulated annealing is a stochastic optimization method inspired by the physical annealing process in metallurgy, where a material is slowly cooled to reach a low-energy crystalline state\cite{kirkpatrick1983optimization}. Unlike population-based algorithms, SA improves a single candidate solution by iteratively exploring neighboring solutions. At each iteration, a new candidate solution is generated by applying a small perturbation to the current solution. If the new solution improves the objective value, it is accepted; otherwise, it may still be accepted with a probability determined by the Metropolis acceptance criterion, which depends on a control parameter known as temperature. As the algorithm progresses, the temperature gradually decreases according to a cooling schedule, reducing the probability of accepting worse solutions and allowing the search to focus on promising regions of the solution space. SA is particularly useful for optimization problems with large or complex search spaces because its probabilistic acceptance mechanism helps with escaping local minima.
\subsection{Performance Comparison: Optimization using BQPhy Q-NOW QIEO vs. Optimization using Traditional Genetic Algorithm and Simulated Annealing on a Classical Computer}

QIEO leverages quantum-inspired evolutionary operations to directly search high-dimensional discrete design spaces, thereby eliminating the need for an explicit surrogate model. In contrast, the classical surrogate approach benefits from reduced evaluation costs during optimization but requires a significant initial investment in training time. This section provides a comparative analysis of the computational cost associated with each approach.

All classical experiments were conducted on a high-performance cluster equipped with an AMD EPYC 7742 chip, featuring 64 cores and 128 threads with a base clock speed of 2.3 GHz and a turbo speed of 3.4 GHz. The system was outfitted with 16 GB of  RAM and an 11.7 TB solid-state drive (SSD). These specifications provide a representative baseline for evaluating computational performance on modern consumer-grade hardware. The second optimization problem given in Eq. \ref{eq:second_prob} is solved with GA, SA and BQPhy Q-NOW QIEO for different domains. All three algorithms, SA \cite{laarhoven1987simulated,kirkpatrick1983optimization}, GA \cite{goldberg1989genetic}, and QIEO \cite{kumar2024benchmarking} were executed using parameter settings suggested in previous studies and commonly adopted in the literature to ensure a fair and balanced comparison. The hyperparameter values used in this study are provided in Table \ref{table:hyperparameters}. The total computational times and optimum solutions for each method are summarized in Table~\ref{tab:comparison}.   
\begin{table}[h!]
\caption{Hyperparameter Settings Used for Optimization Algorithms}
\begin{tabular}{lcc}
\textbf{Method} & \textbf{Parameter} & \textbf{Value} \\
\hline
\multirow{4}{*}{BQPhy Q-NOW QIEO} 
& Population Size & 20 \\
& Rotation Angle & 0.45 \\
& Maximum Iterations & 100 \\
& Function Tolerance & $10^{-6}$ \\
\hline
\multirow{6}{*}{GA} 
& Population Size & 200 \\
& Crossover Rate & 0.8 \\
& Mutation Function & Gaussian \\
& Maximum Stall Generations & 50 \\
& Function Tolerance & $10^{-6}$ \\
& Maximum Generations & 100 \\
& Elite Count & 10 \\
\hline
\multirow{8}{*}{SA} 
& Initial Temperature & 100 \\
& Cooling Schedule & Exponential \\
& Reannealing Interval & 100 \\
& Temperature Function  & Exponential \\
& Annealing Function  & Fast Annealing \\
& Acceptance Function  & Metropolis \\
& Maximum iterations & 10000 \\
& Function Tolerance & $10^{-6}$ \\
\end{tabular}
\label{table:hyperparameters}
\end{table}

\begin{figure}[h!]
\centering
\begin{subfigure}[t]{0.15\textwidth}
    \includegraphics[width=\textwidth]{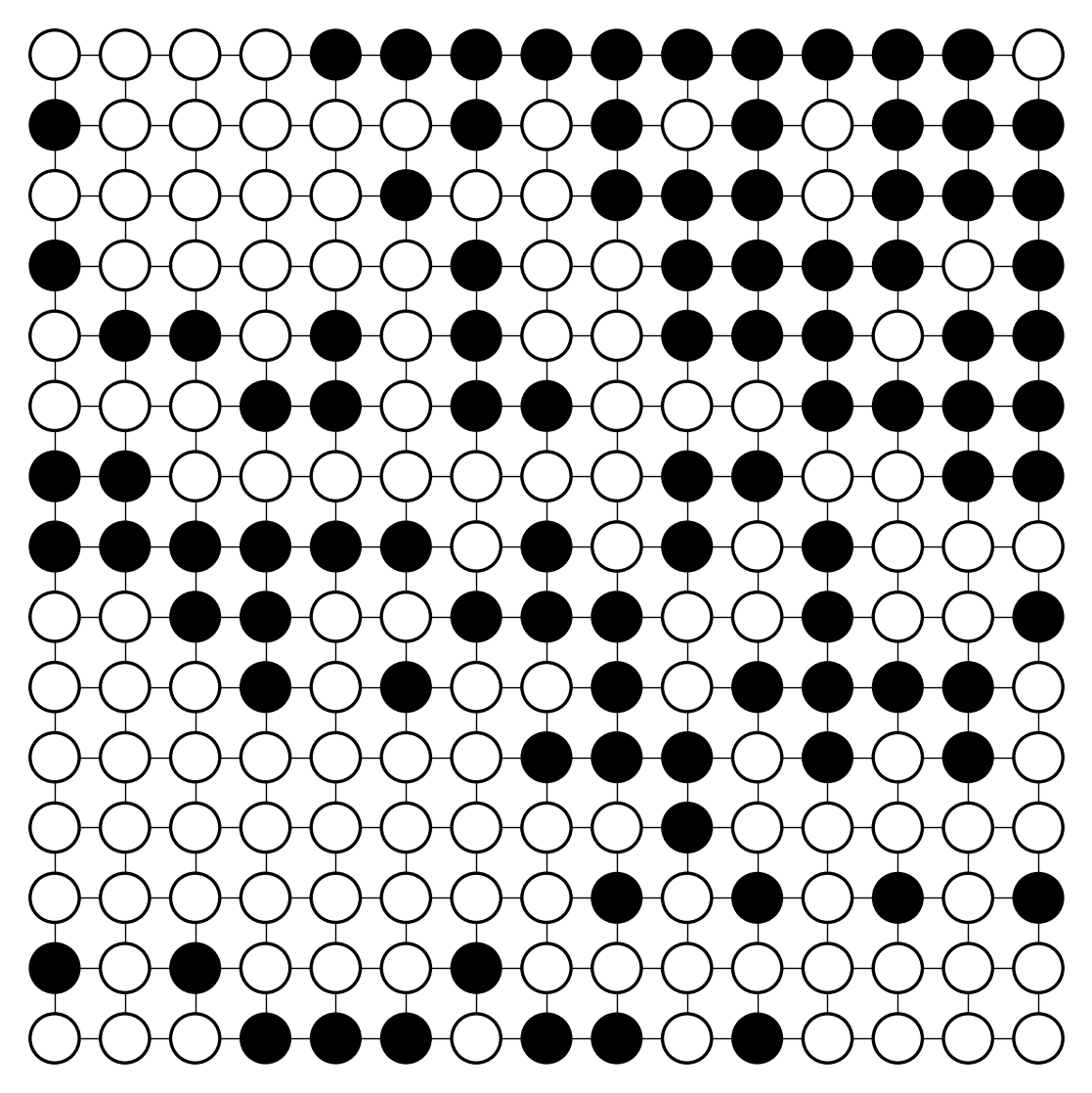}
    \caption{{QIEO Solution} for a $15 \times 15$ Domain}
\end{subfigure}\hspace{\fill} 
\begin{subfigure}[t]{0.15\textwidth}
    \includegraphics[width=\linewidth]{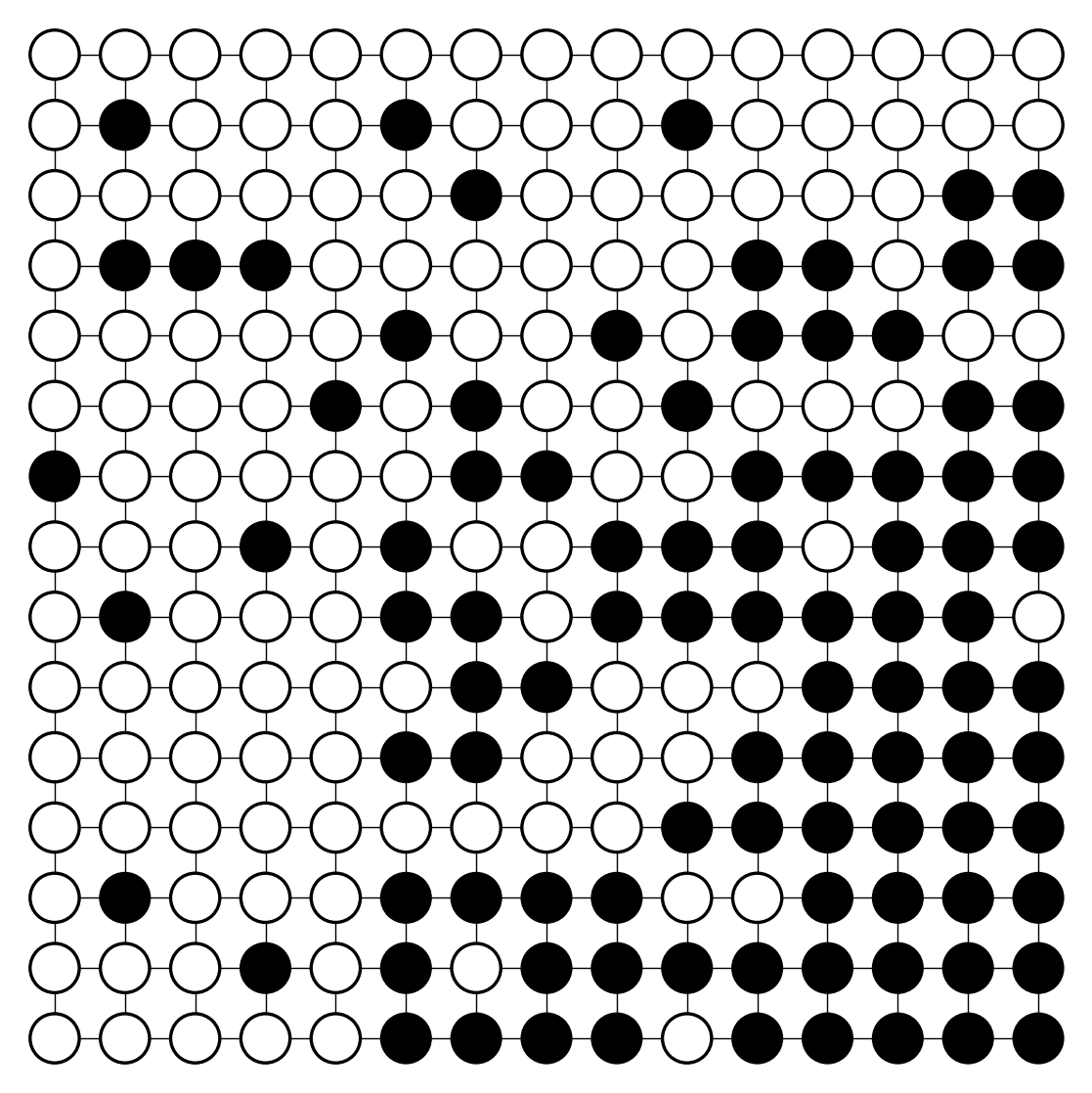}
    \caption{GA Solution for a $15 \times 15$ Domain}
\end{subfigure}\hspace{\fill} 
\begin{subfigure}[t]{0.15\textwidth}
    \includegraphics[width=\linewidth]{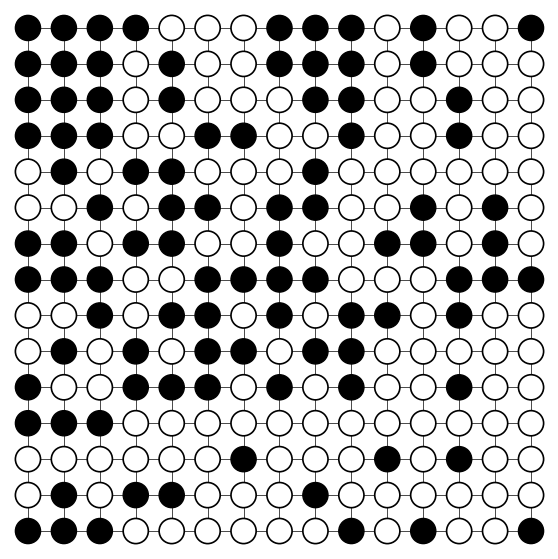}
    \caption{SA Solution for a $15 \times 15$ Domain}
\end{subfigure}

\begin{subfigure}[t]{0.15\textwidth}
    \includegraphics[width=\textwidth]{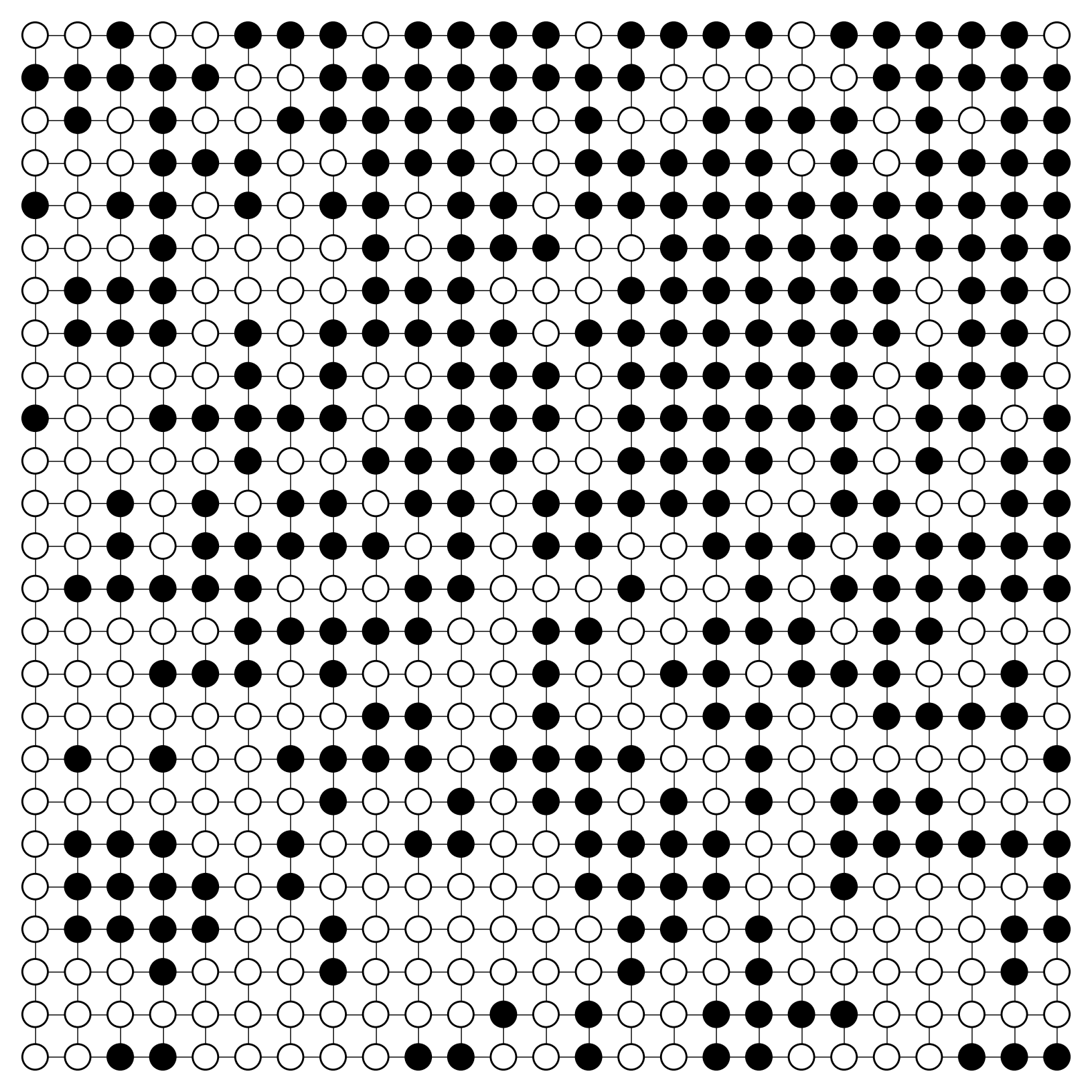}
    \caption{{QIEO Solution} for a $25 \times 25$ Domain}
\end{subfigure}\hspace{\fill} 
\begin{subfigure}[t]{0.15\textwidth}
    \includegraphics[width=\linewidth]{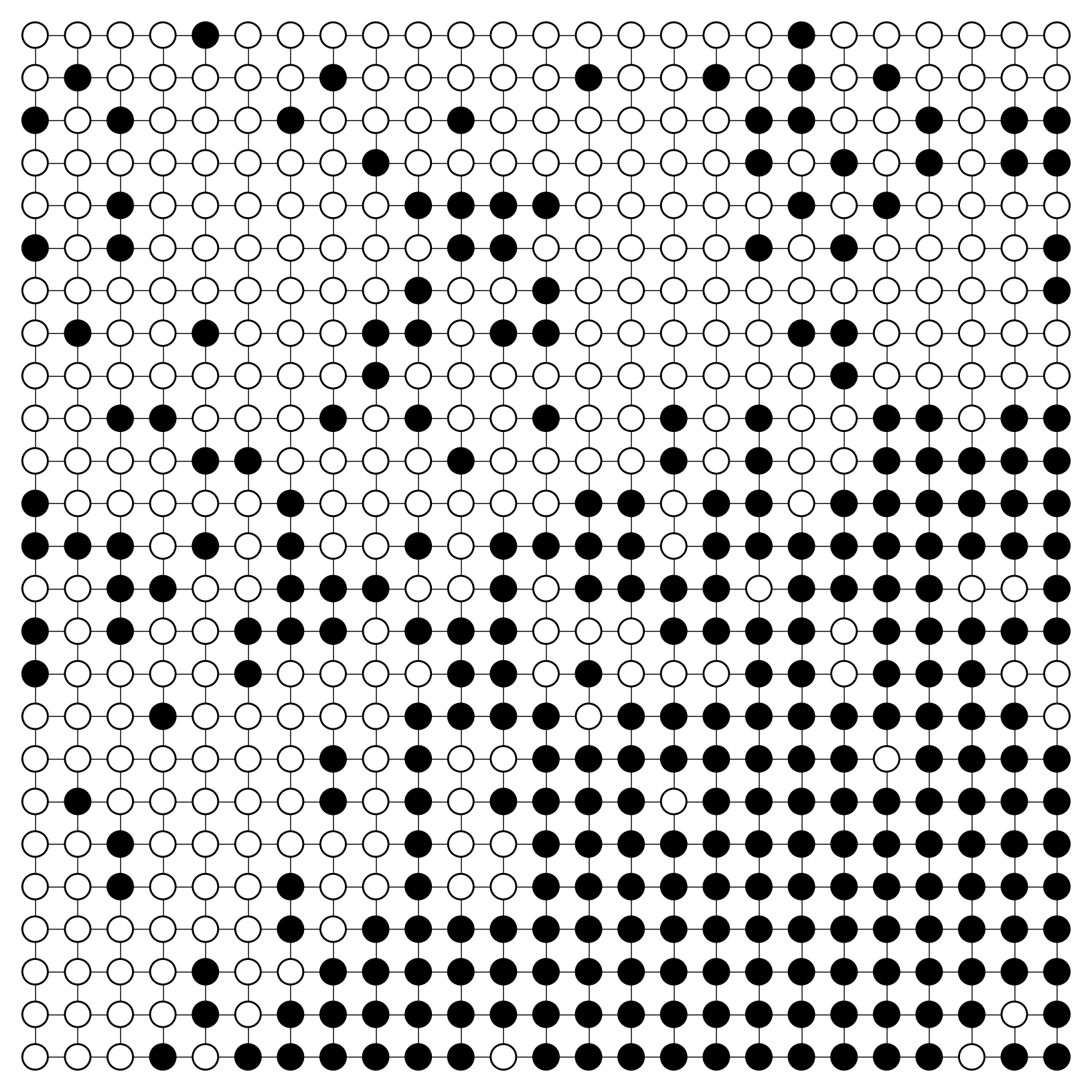}
    \caption{GA Solution for a $25 \times 25$ Domain}
\end{subfigure}\hspace{\fill} 
\begin{subfigure}[t]{0.15\textwidth}
    \includegraphics[width=\linewidth]{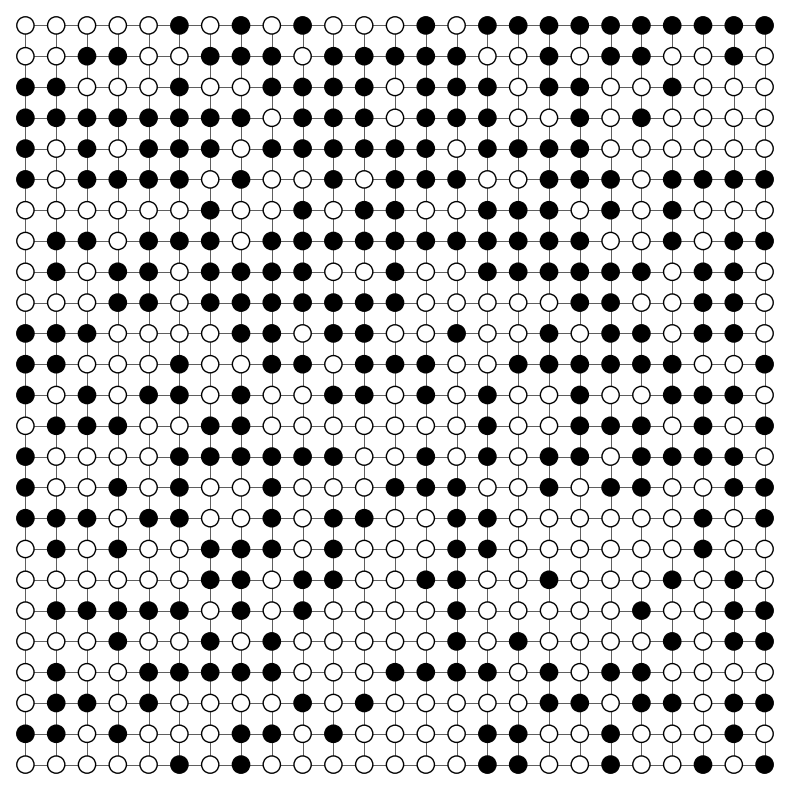}
    \caption{SA Solution for a $25 \times 25$ Domain}
\end{subfigure}

\centering
\begin{subfigure}[t]{0.15\textwidth}
    \includegraphics[width=\textwidth]{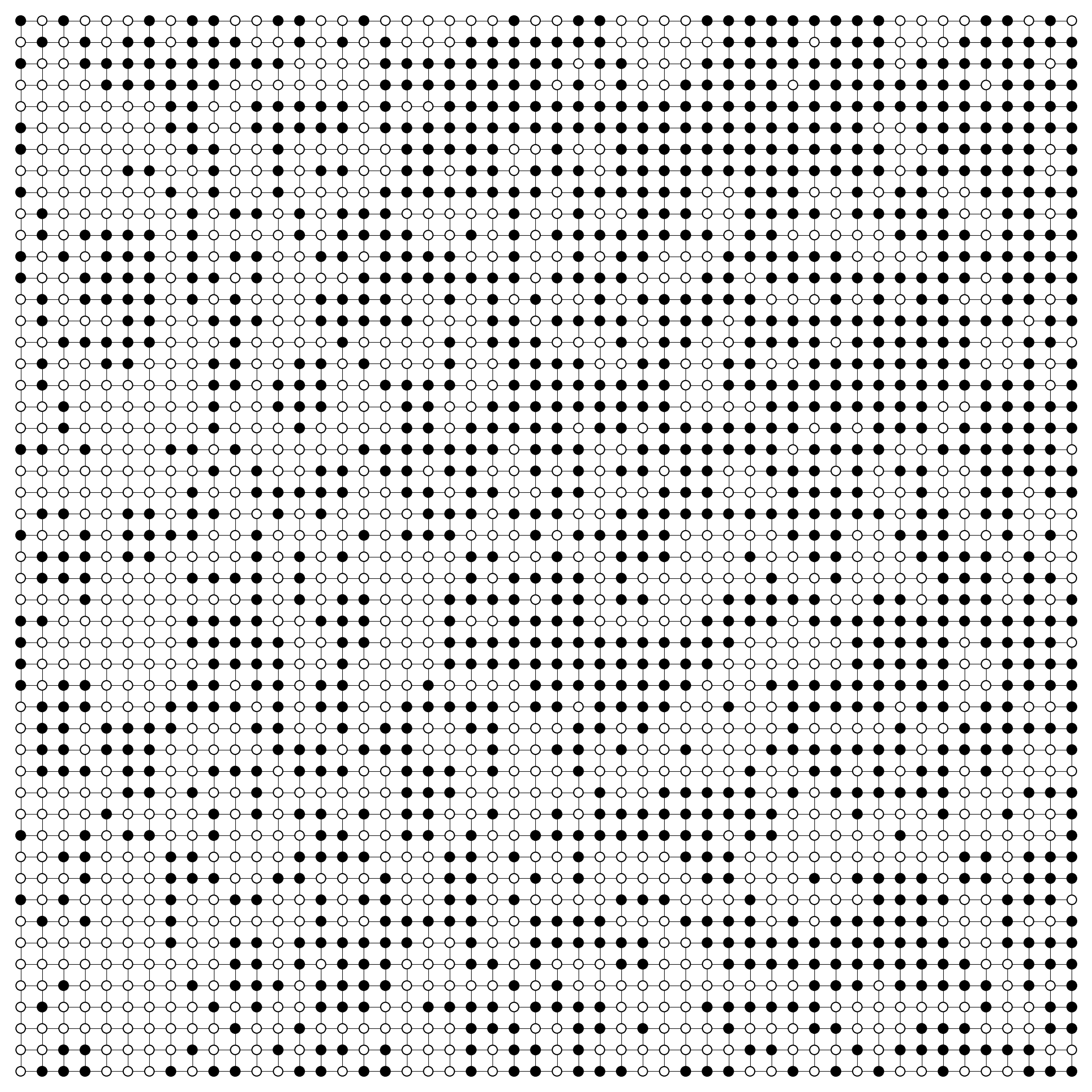}
    \caption{{QIEO Solution} for a $50 \times 50$ Domain}
\end{subfigure}\hspace{\fill} 
\begin{subfigure}[t]{0.15\textwidth}
    \includegraphics[width=\linewidth]{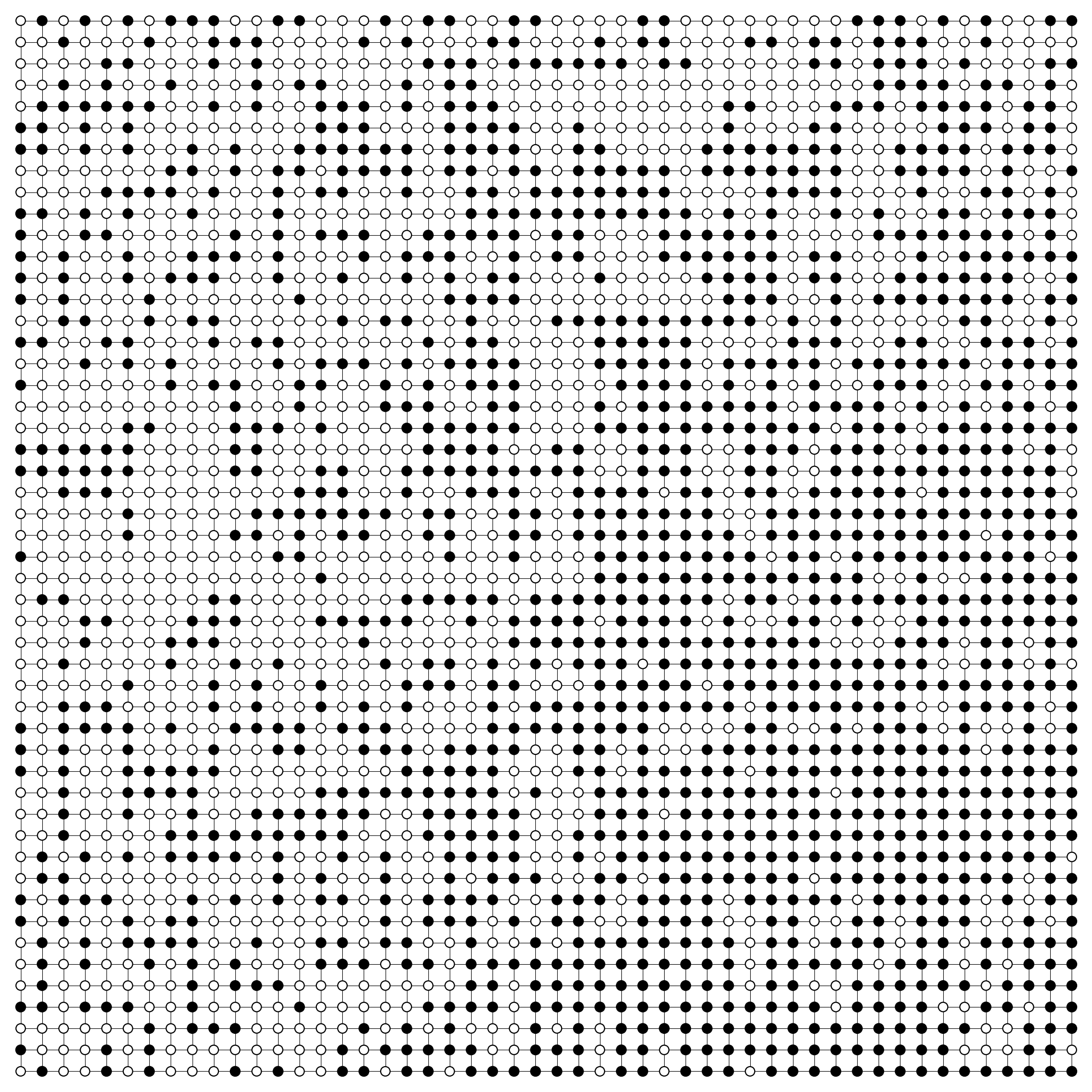}
    \caption{GA Solution for a $50 \times 50$ Domain}
\end{subfigure}\hspace{\fill} 
\begin{subfigure}[t]{0.15\textwidth}
    \includegraphics[width=\linewidth]{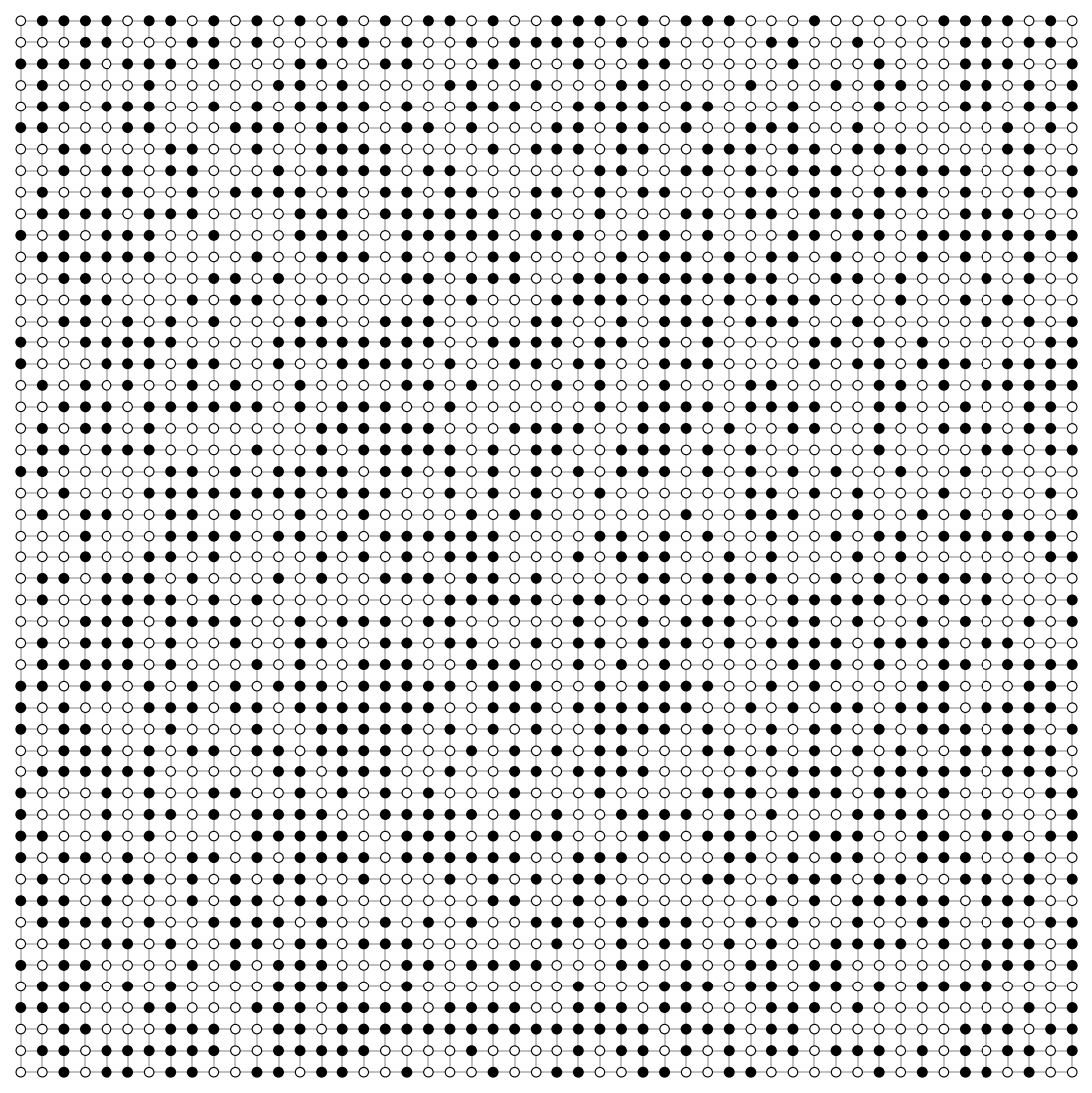}
    \caption{SA Solution for a $50 \times 50$ Domain}
\end{subfigure}
\caption{For comparison, the optimum solutions obtained using QIEO, GA and SA are illustrated for (a, b, c): $15 \times 15$, (d, e, f): $25 \times 25$, and (g, h, i): $50 \times 50$ domains.}
\label{fig:problem3_opt}
\end{figure}

For the traditional genetic algorithm, the optimization required 164.05 seconds for a $10 \times 10$ lattice. In contrast, QIEO solver completed the process in only 75.29 seconds without any loss of accuracy. Extending to larger domains, BQPhy Q-NOW QIEO consistently achieved the same accuracy in roughly half the time of GA. The optimum solutions produced by both methods are presented in Fig.~\ref{fig:problem3_opt}. However, SA is prone to getting trapped in local minima and fails to minimize the objective function as effectively as the other methods. Moreover, it suffers from significant computational overhead, taking nearly three and five times longer for 15×15 and 25×25 lattices, respectively, with the overhead increasing for larger lattices. Since the problem explicitly accounts for uncertainty, the existence of multiple equivalent optima is expected and is indeed observed across all solutions. QIEO direct handling of discrete binary variables may be advantageous primarily for large-scale systems with complex energy landscapes.


\begin{table}[h!]
\caption{Performance Comparison of QIEO, GA and SA in Minimizing $std_{f_m}$} 
\begin{tabular}{lccc}
\textbf{Domain} & \textbf{Method} & \textbf{Optimum Solution } & \textbf{Optimization Time (s)} \\
\hline
\multirow{3}{*}{$10\times 10$} & \textbf{QIEO} & \textbf{0.0184008} & \textbf{75.29}  \\
                    & GA         & 0.0184088 & 164.05  \\ & SA         & 0.0185855 & 376.05  \\
\multirow{3}{*}{$15\times 15$} & \textbf{QIEO} & \textbf{0.0096372} & \textbf{298.10}  \\
                    & GA         & 0.0096985 & 565.94  \\ 
                    & SA         & 0.0097663 & 1476.145  \\ 
\multirow{3}{*}{$25\times 25$} & \textbf{QIEO} & \textbf{0.0040946} & \textbf{1911.785} \\
                    & GA         & 0.0040969 & 3452.094 \\
                    & SA         & 0.0041801 & 17464.92 \\
\multirow{3}{*}{$50\times 50$} & \textbf{QIEO} & \textbf{0.0012304}& \textbf{25600.182} \\
                    & GA         & 0.0012368 & 46576.803  \\
                    & SA         & 0.0012616 & 585946.318 \\
\end{tabular}
\label{tab:comparison}
\end{table}





These results highlight the promise of quantum-inspired methods such as BQPhy Q-NOW QIEO for accelerating the design of magnetic lattices. By substantially reducing computational time while slightly improving the solution quality, QIEO demonstrates scalability and efficiency that classical evolutionary approaches struggle to match. This advantage becomes increasingly significant as system size grows, suggesting that quantum-inspired optimization could play a pivotal role in tackling next-generation problems in materials design.

\section{Discussions and Future Work}
This study demonstrates that the incorporation of uncertainty in external parameters such as temperature and magnetic field fundamentally affects the optimization landscape for spin systems. While the first problem converges almost instantly to a uniform spin configuration, the second problem requires a much larger number of iterations to approach an optimum solution. This increase in computational demand arises from the variability introduced by modeling $T$ and $h$ as Gaussian-distributed random variables and from the necessity of exploring a wide design space to capture their effects. As a result, classical algorithms such as GA become computationally expensive, especially when long-range spin interactions are also considered. 

In this context, the application of BQPhy Q-NOW QIEO offers an important alternative. By encoding discrete spin states through probabilistic amplitudes and updating them with quantum-inspired operators, QIEO provides a more efficient search of the energy landscape. The effectiveness of QIEO is evaluated by comparing the results with GA. QIEO is an evolutionary algorithm that utilizes populations to evolve and update them in subsequent generations through the application of a rotation gate, while GA represents the classical counterpart that performs similar population-level updates using crossover and mutation. On the other hand, SA follows a single-trajectory, temperature-controlled stochastic descent governed by the Metropolis acceptance rule, without memory of past solutions or population-based learning. Therefore, SA belongs to a different optimization paradigm, whereas GA aligns with the population-driven evolutionary nature of QIEO, making GA the most consistent and relevant baseline for comparison. The comparison between GA and QIEO shows that for uncertainty-aware optimization, QIEO achieves comparable objective values at significantly reduced computational times. {This time advantage highlighted in this article is solely due to the algorithmic differences between GA and BQPhy Q-NOW QIEO. The chromosome updating process in QIEO is simpler, specifically theta updation, which involves fewer calculations, whereas GA uses more parameters and requires a relatively higher number of calculations \cite{kumar2024qieo,kumar2025qieo,kumar2025qieoSci}.} 

{Table \ref{tab:comparison} also highlights the accuracy advantage of BQPhy's QIEO over both SA and GA for different lattice structures. The lower optimum value is the better solution, as it represents the optimal solution in a minimization problem. For all cases, the optimum solution is slightly better for QIEO when compared with GA and SA, but the optimal lattices obtained from these are quite different (see Fig. \ref{fig:problem3_opt}).} This efficiency advantage shows the potential of quantum-inspired approaches for complex Hamiltonian-based models where classical methods scale poorly. Moreover, the success of BQPhy Q-NOW QIEO in this study highlights that hybrid strategies, which leverage quantum principles while operating on classical high-performance computing platforms, can serve as an effective interim solution until large-scale quantum hardware becomes more accessible.

Future work will extend the present framework along several directions. First, while the current study focuses on a maximum $50 \times 50$ lattice, larger and higher-dimensional systems remain to be investigated, where the computational burden of GA may become prohibitive and the benefits of BQPhy Q-NOW QIEO more pronounced. Second, additional forms of uncertainty, such as variations in long-range weight parameters $\omega_{j}$ or lattice geometry, could be incorporated to further evaluate robustness. Third, the role of long-range interactions, modeled via Gaussian distributions, could be explored in more detail under uncertainty to assess their contribution to critical phenomena, such as the ferromagnetic–paramagnetic phase transition. Finally, the comparative performances between GA, SA, and QIEO across these more demanding scenarios will help clarify the domains in which quantum-inspired methods consistently outperform classical optimization.

\section*{Conflict of Interest Statement}
On behalf of all authors, the corresponding author states that there is no conflict of interest.

\section*{Credit}
The following article has been submitted to \textit{APL Quantum}. After it is published, it will be found at https://publishing.aip.org/resources/librarians/products/journals/.

\section*{Acknowledgments}
Virginia Tech authors would like to acknowledge the financial support from the Air Force Office of Scientific Research (AFOSR) Grants FA9550-21-1-0120 and FA9550-24-1-0154 (Program Officer: Dr. Ali Sayir) for the development of the Ising model solution and magnetic domain formulation.

\bibliography{aipsamp}

\end{document}